\documentclass{article}%
\usepackage{amsmath}
\usepackage{amsfonts}
\usepackage{amssymb}
\usepackage{graphicx}%
\providecommand{\U}[1]{\protect\rule{.1in}{.1in}}
\newtheorem{theorem}{Theorem}[section]

\newtheorem{corollary}[theorem]{Corollary}

\newtheorem{definition}[theorem]{Definition}
\newtheorem{example}[theorem]{Example}

\newtheorem{lemma}[theorem]{Lemma}

\begin{document}

\title{Dual method for continuous-time Markowitz's Problems with nonlinear wealth equations}
\author{Shaolin Ji \thanks{School of Mathematics \& System Sciences, Shandong
University, Jinan 250100, People's Republic of China. email: jsl@sdu.edu.cn.
This work is supported by the National Basic Research Program of China (973
Program, No. 2007CB814900). This work was submitted to SIAM Journal on Control
and Optimization on August 21, 2007 (manuscript number: 070066).} }
\date{}
\maketitle

\textbf{Abstract. }Continuous-time mean-variance portfolio selection model
with nonlinear wealth equations and bankruptcy prohibition is investigated by
the dual method. A necessary and sufficient condition which the optimal
terminal wealth satisfies is obtained through a terminal perturbation
technique. It is also shown that the optimal wealth and portfolio is the
solution of a forward-backward stochastic differential equation with constraints.

\vskip0.2cm


\vskip0.2cm


\baselineskip=16pt

\pagestyle{myheadings} \thispagestyle{plain}

\section{Introduction\-}

Mean-variance portfolio selection in discrete time setting has been well
studied. But mean-variance portfolio selection has received little attention
in the context of continuous-time models \cite{st}. Recently several papers
studied various continuous-time Markowitz's models [2, 12-16, 27, 28]. There
are mainly two approaches which are employed to study this problem in
continuous-time case: the forward (primal) method [15, 16, 27] which is
inspired by the indefinite LQ control theory \cite{yo-zh}, and backward (dual)
method which is employed by Bielecki et al. \cite{bi-ji}.

The dual method (also known as martingale method) is first studied by Harrison
and Kreps \cite{ha-kr} and Pliska [23, 24]. A systematic account on this
method and its application to utility optimization problems can be found in
\cite{ka} and the references therein. It mainly includes two steps: the first
step is to compute the optimal terminal wealth, and the second one is to
compute the portfolio strategy replicating the obtained optimal terminal
wealth. It is worth pointing out that the dual method is powerful in solving
stochastic control problem with sample-wise constraint imposed on the state. A
sample-wise constraint requires that the state be in a given set with
probability 1; for example, a nonnegativity constraint on the wealth process,
i.e., bankruptcy prohibition. For a deeper discussion we refer the reader to a
recent paper by Ji and Zhou \cite{ji-zhou}.

In this paper, we study the continuous-time mean-variance portfolio selection
model with nonlinear wealth equation and bankruptcy prohibition. To apply the
dual method, we first give a backward formulation of this problem in which the
terminal wealth is regarded as the \textquotedblleft control variable". Note
that, in this formulation, the initial wealth becomes an additional
constraint. Under convexity assumptions, the backward formulation leads to a
static convex programming problem. Then a terminal perturbation technique is
introduced to derive a stochastic maximum principle which characterizes the
optimal terminal wealth. Due to the convexity assumptions on the coefficients,
we prove that the established stochastic maximum principle is also a
sufficient condition. The terminal perturbation technique is first studied in
El Karoui, Peng and Quenez \cite{el-pe-2001} to solve a recursive utility
optimization problem. Recently, Ji and Peng \cite{ji-peng} use this technique
and Ekeland's variational principle to obtain a necessary condition for the
mean-variance portfolio selection problem with non-convex wealth equations.
Finally, we show that the optimal wealth and portfolio can be solved by a
forward-backward stochastic differential equation (FBSDE) with constraints.

This paper is organized as follows: in section 2, we introduce continuous-time
mean-variance portfolio selection model with nonlinear wealth equation and
bankruptcy prohibition as well as its equivalent backward formulation.
Applying Lagrange multiplier and terminal perturbation technique, we obtain a
necessary and sufficient condition for optimality in section 3. In section 4,
we prove that there exists an optimal solution of the continuous-time
mean-variance portfolio selection problem and it can be obtained by solving a
FBSDE. Finally, section 5 closes the paper with some concluding remarks.

\section{Problem formulation}

Let $W(\cdot)=(W_{1}(\cdot),\ldots,W_{d}(\cdot))^{\prime}$ be a standard
d-dimensional Brownian Motion defined on a complete probability space
$(\Omega,\mathcal{F},P)$. The information structure is given by a filtration
$F=\{\mathcal{F}_{t}\}_{0\leq t\leq T}$, which is the $\sigma-$algebra
generated by the Brownian Motion $W(\cdot)$ and augmented. For any given
Euclidean space $H$, we denote by $M^{2}(0,T;H)$, the space of all
$\mathcal{F}_{t}-$progressively measurable processes $x(\cdot)$\ with values
in $H,$ such that%
\[
E\int_{0}^{T}\mid x(t)\mid^{2}dt<\infty.
\]
Denote by $L^{2}(\Omega,\mathcal{F}_{T},P)$, the space of all $\mathcal{F}%
_{T}-$measurable random variable $\xi$ with value in $R$, such that $E\mid
\xi\mid^{2}<\infty$.

\subsection{The wealth process}

Consider a complete market where there are one bank account (risk free
instrument) and $d$ stocks (risky instruments), and an investor who can decide
at time $t\in\lbrack0,T]$ the amount $\pi_{i}(t)$ to invest in the $i$th stock
($i=1,\ldots d$) with initial investment $x>0$. The respective prices of the
instruments are $S_{0}(\cdot)$ and $S_{1}(\cdot),\cdots,S_{d}(\cdot)$, and the
portfolio is $\pi(\cdot)=(\pi_{1}(t),\ldots,\pi_{d}(t))^{\prime}.$ We suppose
that the wealth process $X(\cdot)$ is governed by the following stochastic
differential equation%
\begin{equation}
\left\{
\begin{array}
[c]{l}%
-dX(t)=f(X(t),\sigma(t)^{\prime}\pi(t),t)dt-\pi(t)^{\prime}\sigma(t)dW(t),\\
X(0)=x
\end{array}
\right.  \tag{2.1}%
\end{equation}
where the stock-volatility matrix $\sigma(\cdot)=\{\sigma_{ij}(\cdot)\}_{1\leq
i,j\leq d}$ is a predictable and bounded process. $\sigma(\cdot)$ is also
assumed to be invertible and $\sigma^{-1}(\cdot)$ be bounded uniformly in
$(t,\omega)\in\lbrack0,T]\times\Omega$. Set $Z(t)=\sigma(t)^{\prime}\pi(t).$
Then (2.1) can be rewritten as%
\begin{equation}
\left\{
\begin{array}
[c]{l}%
-dX(t)=f(X(t),Z(t),t)dt-Z(t)^{\prime}dW(t),\\
X(0)=x.
\end{array}
\right.  \tag{2.2}%
\end{equation}

We assume

(H1) $f$ is continuous in $R\times R^{d}\times\lbrack0,T]$ for $a.a.\omega$
and has continuous bounded derivatives in $(X,Z)$;

(H2) $f(0,0,\cdot,\cdot)\in M^{2}(0,T;R);$

(H3) $f$ is convex with respect to $(X,Z);$

(H4) $f(0,0,t)\geq0$ a.s.

In the following, we give two specific examples to illustrate the model (2.1).

\begin{example}
The standard linear case.

The prices $S_{0}(\cdot)$ and $S_{1}(\cdot),\cdots,S_{d}(\cdot)$ are governed
by the equations%
\[%
\begin{array}
[c]{ll}%
dS_{0}(t)=S_{0}(t)r(t)dt, & S_{0}(0)=s_{0};\\
dS_{i}(t)=S_{i}(t)[b_{i}(t)dt+\sum\limits_{j=1}^{d}\sigma_{ij}(t)dW(t)], &
S_{i}(0)=s_{i}>0;i=1,\ldots,d.
\end{array}
\]

We assume: the interest rate $r(\cdot)$ is a non-negative, predictable and
uniformly bounded scalar-valued process; the stock-appreciation rates
$b(\cdot)=(b_{1}(\cdot),\ldots b_{d}(\cdot))^{\prime}$ is a predictable and
uniformly bounded process.

Set $B(t):=(b_{1}(t)-r(t),\ldots,b_{m}(t)-r(t))^{\prime}$. Define the risk
premium process $\theta(t)\equiv(\theta_{1}(t),\ldots,\theta_{m}(t))^{\prime
}:=\sigma(t)^{-1}B(t)$. The wealth process $X(\cdot)$ satisfies the following
linear stochastic differential equation%
\begin{equation}
\left\{
\begin{array}
[c]{l}%
dX(t)=[r(t)X(t)+\pi(t)^{\prime}\sigma(t)\theta(t)]dt+\pi(t)^{\prime}%
\sigma(t)dW(t),\\
X(0)=x.
\end{array}
\right.  \tag{2.3}%
\end{equation}
Note that for this case,%
\[
f(X,\sigma(t)^{\prime}\pi,t)=-r(t)X-\pi^{\prime}\sigma(t)\theta(t).
\]

\end{example}

\begin{example}
A large investor case.

An interesting example of a nonlinear wealth equation is the optimal portfolio
choice problem for a large investor considered in Cuoco and Cvitanic
\cite{cu-cv}. Refer to [3, 5, 7]\ for other models. In \cite{cu-cv},
$S_{0}(\cdot)$ and $S_{1}(\cdot),\cdots,S_{d}(\cdot)$ are described by
equations%
\[%
\begin{array}
[c]{ll}%
dS_{0}(t)=S_{0}(t)[r(t)+l_{0}(X(t),\pi(t))]dt, & S_{0}(0)=s_{0};\\
dS_{i}(t)=S_{i}(t)[(b_{i}(t)+l_{i}(X(t),\pi(t)))dt+\sum\limits_{j=1}^{d}%
\sigma_{ij}(t)dW(t)], & S_{i}(0)=s_{i}>0;i=1,\ldots,d
\end{array}
\]
where $l_{i}:R^{+}\times R^{d}\rightarrow R,\,0\leq i\leq d$ are given
functions which describe the effect of the wealth and the strategy. In this
case,%
\[
f(X,\sigma(t)^{\prime}\pi,t)=-r(t)X-(X-\pi^{^{\prime}}\mathbf{1)}l_{0}%
(X,\pi)-\pi^{^{\prime}}[b(t)-r(t)\mathbf{1}+l(X,\pi)].
\]

\end{example}

\subsection{Backward formulation of the problem}

Before formulating the problem, we point out that we distinguish the concepts
between initial investment and initial wealth. Throughout this paper, we
suppose that the initial investment $x$ of the investor is less than or equal
to his initial wealth $y$, i.e. $x\leq y$.

Usually the continuous-time mean-variance portfolio selection problem with
bankruptcy prohibition is formulated as: the investor chooses his portfolio
and initial investment $x$ so as to%
\[
Minimize\qquad Var\,X(T)\equiv EX(T)^{2}-c^{2},
\]%
\begin{equation}
subject\,\,to\left\{
\begin{array}
[c]{l}%
EX(T)=c,\\
X(t)\geq0\qquad a.s.,\quad t\in\lbrack0,T],\\
\pi(\cdot)\in M^{2}(0,T;R^{m}),\\
(X(\cdot),\pi(\cdot))\,\text{satisfies\thinspace equation }(2.1)\text{ and
}0<x=X(0)\leq y,
\end{array}
\right.  \tag{2.4}%
\end{equation}
where $c>0$ is a given expectation level with respect to the investor's
terminal wealth $X(T)$, and $X(t)\geq0$ means that no-bankruptcy is required.

\begin{definition}
A portfolio $\pi(\cdot)$ is said to be admissible if $\pi(\cdot)\in
M_{\mathcal{F}}^{2}(0,T;R^{m}),$ $EX(T)=c$ and the corresponding wealth
processes $X(t)\geq0\,a.s,\forall\,t\in\lbrack0,T].$
\end{definition}

We denote by $\mathcal{A}(x)$ the set of portfolio $\pi(\cdot)$ admissible for
the initial investment $x$. Set
\begin{equation}
V(y)=\underset{0<x\leq y,\pi\in\mathcal{A}(x)}{\min}\{E[X^{x,\pi}%
(T)]^{2}-c^{2}\}. \tag{2.5}%
\end{equation}

In the following we give an equivalent backward formulation of the above
optimization problem (2.4).

Since $\sigma(\cdot)$ is invertible, $Z(\cdot)$ can be regarded as the
"control variable" instead of $\pi(\cdot)$. Notice that selecting $Z(\cdot
)$\ is equivalent to selecting the terminal wealth $X(T)$ by the backward
stochastic differential equation (BSDE) theory \cite{pa-pe}. Hence the wealth
equation (2.2) can be rewritten as
\begin{equation}
\left\{
\begin{array}
[c]{l}%
-dX(t)=f(X(t),Z(t),t)dt-Z(t)^{\prime}dW(t),\\
X(T)=\xi
\end{array}
\right.  \tag{2.6}%
\end{equation}
where the terminal wealth $\xi$\ is the "control" to be chosen from the
following set%
\[
U=\{\xi\mid\xi\in L^{2}(\Omega,\mathcal{F}_{T},P),\xi\geq0,a.s.\}.
\]
Note that nonnegative terminal wealth, i.e., $\xi=x(T)\geq0$ keeps the wealth
process nonnegative all the time, as implied by Assumption (H4) and the
comparison theorem for BSDEs.

This gives rise to the following optimization problem:%
\[
\text{Minimize}\qquad J(\xi)\triangleq(E\xi^{2}-c^{2})
\]%
\begin{equation}
subject\,\,to\left\{
\begin{array}
[c]{l}%
E\xi=c,\\
X(0)\leq y,\\
\xi\in U.
\end{array}
\right.  \tag{2.7}%
\end{equation}

It is clear that the original problem (2.4) is equivalent to (2.7). Hence,
hereafter we focus ourselves on solving (2.7). The advantage of doing this
lies in the fact that the state constraint in (2.4) now becomes a control
constraint in (2.7) since $\xi$ is regarded as the control variable. It is
well known in control theory that a control constraint is easier to deal with
than a state constraint. But there is a cost of doing so: the original initial
condition $X(0)=x$ now becomes a constraint, i.e., $X(0)\leq y$.

It is easy to prove that Assumptions (H1) and (H2) ensure there exists a
unique pair $(X(\cdot),Z(\cdot))\in M^{2}(0,T;R)\times M^{2}(0,T;R^{d})$ of
(2.6) \cite{pa-pe}. From now on, we denote the solution of (2.6) by $(X^{\xi
}(\cdot),Z^{\xi}(\cdot))$, whenever necessary, to show the dependence on $\xi
$. We also denote $X^{\xi}(0)$ by $X_{0}^{\xi}$.

\begin{definition}
$\xi$ is called admissible for given $y>0$ and $c>0$, if $\xi\in U$ and the
solution of (2.6) satisfies $X_{0}^{\xi}\leq y,$ $E\xi=c$. We shall denote by
$\mathcal{N}(y),$ the set of all admissible $\xi^{\prime}$s for any given $y$
and $c$.
\end{definition}

An admissible $\xi^{\ast}$ is called optimal if it attains the minimum of
$J(\xi)$ over $\mathcal{N}(y)$. From above discussions, we know that
$V(y)=J(\xi^{\ast})$. The optimal portfolio for (2.7) is called a variance
minimizing\thinspace\thinspace portfolio.\ After the optimal terminal wealth
$\xi^{\ast}$ is obtained, we can compute the optimal portfolio by solving (2.6).

For the feasibility of above optimization problem (2.4) and (2.7), we assume
the following slater condition:

(H5) For given $y>0$ and $c>0$, there exist an initial investment $x^{o}$
($0<x^{o}<y$) and a portfolio $\pi^{o}$ such that the corresponding terminal
wealth $X^{o}(T)\geq0$ and $EX^{o}(T)=c$.

\textbf{Remark}. In fact, the feasibility of (2.4) and (2.7) can be checked by
solving another optimization problem. For more details, see Appendix A.

Note that if $y\geq X_{0}^{c}$, then $\xi\equiv c$ is admissible. In this
case, it is obvious that $V(y)=0$. Hence, without loss of generality we can assume

(H6) $y<X_{0}^{c}.$

\section{A sufficient and necessary condition for optimality}

In this section, we derive a sufficient and necessary condition which
characterizes the optimal terminal wealth.

It is easy to check that the following $R-$valued functionals on $U$%
\begin{align*}
\xi &  \mapsto X_{0}^{\xi}-y,\\
\xi &  \mapsto E\xi^{2}-c^{2},\\
\xi &  \mapsto E\xi-c
\end{align*}
are convex under Assumption (H3). Hence, applying classical results of convex
analysis \cite{lu}, it is easy to obtain the following lemma.

\begin{lemma}
We suppose (H1)-(H6). There exist real numbers $\lambda_{1}\geq0$ and
$\lambda_{2}$ such that%
\begin{equation}
V(y)=\underset{\xi\in U}{\min}\{E\xi^{2}-c^{2}+\lambda_{1}(X_{0}^{\xi
}-y)+\lambda_{2}(E\xi-c)\}. \tag{3.1}%
\end{equation}
Furthermore, if the minimum is attained in (2.7) by $\xi^{\ast}$, then it is
attained in (3.1) by $\xi^{\ast}$ with $\lambda_{1}(X_{0}^{\xi^{\ast}}-y)=0.$
Conversely, suppose there exist $\lambda_{1}^{o}\geq0$, $\lambda_{2}^{o}\in R$
and $\xi^{o}\in U$ such that the minimum is achieved in%
\[
\underset{\xi\in U}{\min}\{E\xi^{2}-c^{2}+\lambda_{1}^{o}(X_{0}^{\xi
}-y)+\lambda_{2}^{o}(E\xi-c)\}
\]
with $\lambda_{1}^{o}(X_{0}^{\xi^{o}}-y)=0$, then the minimum is achieved in
(2.7) by $\xi^{o}$.
\end{lemma}

In the following, we introduce a terminal perturbation technique which is used
in [7, 10].

Let $\xi^{\ast}$ be optimal for (2.7) and $(X^{\ast}(\cdot),Z^{\ast}(\cdot))$
be the corresponding optimal trajectory, i.e., the solution of (2.6) under
$\xi^{\ast}$. Let $\hat{\xi}\in L^{2}(\Omega,\mathcal{F},P)$ such that
$(\xi^{\ast}+\hat{\xi})\in U$. Since $U$ is convex, then for any $0\leq
\rho\leq1$,%
\[
\xi^{\rho}\triangleq\xi^{\ast}+\rho\hat{\xi}%
\]
is also in $U$. Let $(\delta X(\cdot),\delta Z(\cdot))$ be the solution of the
following first order variational equation%
\begin{equation}
\left\{
\begin{array}
[c]{l}%
-d\delta X(t)=[f_{X}(X^{\ast}(t),Z^{\ast}(t),t)\delta X(t)+f_{Z}(X^{\ast
}(t),Z^{\ast}(t),t)\delta Z(t)]dt-\delta Z(t)^{\prime}dW(t),\\
\delta X(T)=\hat{\xi}.
\end{array}
\right.  \tag{3.2}%
\end{equation}
Note that (3.2) is a linear BSDE and it has a unique pair $(\delta
X(\cdot),\delta Z(\cdot))\in M^{2}(0,T;R)\times M^{2}(0,T;R^{d})$. We denote
by $(X^{\rho}(\cdot),Z^{\rho}(\cdot))$ the solution of (2.6) corresponding to
$X(T)=\xi^{\rho}$. Set%
\begin{align*}
\tilde{X}^{\rho}(t)  &  =\rho^{-1}[X^{\rho}(t)-X^{\ast}(t)]-\delta X(t),\\
\tilde{Z}^{\rho}(t)  &  =\rho^{-1}[Z^{\rho}(t)-Z^{\ast}(t)]-\delta Z(t).
\end{align*}
Using the techniques in \cite{peng-1993}, we have the following convergence results.

\begin{lemma}
Assume (H1) and (H2), then%
\begin{align*}
\underset{\rho\rightarrow0}{\lim}\underset{0\leq t\leq T}{\sup}E  &
\mid\tilde{X}^{\rho}(t)\mid^{2}=0,\\
\underset{\rho\rightarrow0}{\lim}E\int_{0}^{T}  &  \mid\tilde{Z}^{\rho}%
(t)\mid^{2}dt=0.
\end{align*}

\end{lemma}

For the reader's convenience, we sketch the proof of Lemma 3.2 in the Appendix B.

In order to derive the necessary condition, we introduce the adjoint equation%
\begin{equation}
\left\{
\begin{array}
[c]{l}%
dq(t)=q(t)[f_{X}(X^{\ast}(t),Z^{\ast}(t),t)dt+f_{Z}(X^{\ast}(t),Z^{\ast
}(t),t)^{\prime}dW(t)],\\
q(0)=1
\end{array}
\right.  \tag{3.3}%
\end{equation}
where $(X^{\ast}(\cdot),Z^{\ast}(\cdot))$ is the optimal trajectory with
respect to $\xi^{\ast}$. (3.3) is a linear stochastic differential equation
and it has a unique solution in $M^{2}(0,T;R).$

Set%
\[
M\triangleq\{\omega\in\Omega\mid\xi^{\ast}(\omega)=0\}.
\]

\begin{theorem}
We assume (H1)-(H6). $\xi^{\ast}$ is optimal to (2.7) if and only if there
exist constants $\lambda_{1}>0$ and $\lambda_{2}\in R$ such that%
\begin{equation}%
\begin{array}
[c]{l}%
2\xi^{\ast}(\omega)+\lambda_{1}q_{T}(\omega)+\lambda_{2}\geq0\text{ a.s.\quad
on }M,\\
2\xi^{\ast}(\omega)+\lambda_{1}q_{T}(\omega)+\lambda_{2}=0\text{ a.s.\quad on
}M^{c}%
\end{array}
\tag{3.4}%
\end{equation}
with $X_{0}^{\xi^{\ast}}=y,$ where $q(t)$ is the solution of the adjoint
equation (3.3).
\end{theorem}

\textbf{Proof.} (1) Proof of the necessary condition.

By Lemma 3.1, there exist constants $\lambda_{1}\geq0$ and $\lambda_{2}$ such
that%
\[
E(\xi^{\rho})^{2}-c^{2}+\lambda_{1}(X_{0}^{\xi^{\rho}}-y)+\lambda_{2}%
(E\xi^{\rho}-c)\geq E(\xi^{\ast})^{2}-c^{2}+\lambda_{1}(X_{0}^{\xi^{\ast}%
}-y)+\lambda_{2}(E\xi^{\ast}-c).
\]

Dividing the inequality by $\rho$ and sending $\rho$ to $0$, we obtain%
\begin{equation}
2E(\xi^{\ast}\hat{\xi})+\lambda_{1}\delta X(0)+\lambda_{2}E\hat{\xi}\geq0
\tag{3.5}%
\end{equation}
where $\delta X(0)$ denotes the solution of (3.2) at time $0$.

Applying It\^{o}'s lemma to $\delta X(t)q(t)$ yields%
\[%
\begin{array}
[c]{ll}
& E[\delta X(T)\cdot q(T)-\delta X_{0}\cdot q(0)]\\
= & E[-\int_{0}^{T}[(f_{X}(X^{\ast}(t),Z^{\ast}(t),t)\delta X(t)+f_{Z}%
^{\prime}(X^{\ast}(t),Z^{\ast}(t),t)\delta Z(t))q(t)]dt+\\
& \int_{0}^{T}[(f_{X}(X^{\ast}(t),Z^{\ast}(t),t)\delta X(t)q(t)+<\delta
Z(t),f_{Z}(X^{\ast}(t)(t),Z^{\ast}(t),t)q(t)>)]dt\\
= & 0.
\end{array}
\]

Since $q(0)=1$, it is obvious that%
\begin{equation}
\delta X_{0}=E[\hat{\xi}\cdot q(T)]. \tag{3.6}%
\end{equation}

Replacing $\delta X_{0}$ with $E[\hat{\xi}\cdot q(T)]$\ in (3.5), we have that
for each $\bar{\xi}\in U$, the following inequality holds%
\begin{align}
&  2E(\xi^{\ast}\hat{\xi})+\lambda_{1}E[\hat{\xi}\cdot q(T)]+\lambda_{2}%
E\hat{\xi}\tag{3.7}\\
&  =E[(2\xi^{\ast}+\lambda_{1}q(T)+\lambda_{2})\cdot\hat{\xi}]\nonumber\\
&  =E[(2\xi^{\ast}+\lambda_{1}q(T)+\lambda_{2})\cdot(\bar{\xi}-\xi^{\ast
})]\nonumber\\
&  \geq0.\nonumber
\end{align}

Thus, it is easy to check that for each $\varepsilon>0$%
\[
P\{\omega\mid\omega\in M,\,2\xi^{\ast}+\lambda_{1}q(T)+\lambda_{2}%
<-\varepsilon\}=0.
\]

From the continuity property of probability, we have%
\[
2\xi^{\ast}+\lambda_{1}q(T)+\lambda_{2}\geq0\text{ a.s.\quad on }M.
\]

By a similar argument,%
\[
2\xi^{\ast}+\lambda_{1}q(T)+\lambda_{2}=0\text{ a.s.\quad on }M^{c}.
\]

Now we show that $\lambda_{1}\neq0.$ If $\lambda_{1}=0$, (3.4) becomes%
\begin{equation}%
\begin{array}
[c]{l}%
\xi^{\ast}(\omega)\geq-\frac{\lambda_{2}}{2}\text{ a.s.\quad on }M,\\
\xi^{\ast}(\omega)=-\frac{\lambda_{2}}{2}\text{ a.s.\quad on }M^{c}.
\end{array}
\tag{3.8}%
\end{equation}

There are two cases: one is $M$ is nonempty and the other is $M$ is empty. For
the first case, we deduce that $\xi^{\ast}=0$ which contradicts to the
constraint $E\xi^{\ast}=c>0$. For the second case, we have that $\xi^{\ast}=c$
from (3.8) and the constraint $E\xi^{\ast}=c$. But this contradicts to
Assumption (H6). In summary, we have $\lambda_{1}>0.$

By Lemma 3.1, we know $\lambda_{1}(X_{0}^{\xi^{\ast}}-y)=0$. Since
$\lambda_{1}>0$, it is easy to see $X_{0}^{\xi^{\ast}}=y$ holds.

(2) Proof of the sufficient condition.

Let $\xi\in U$ with $(X(\cdot),Z(\cdot))$ be the corresponding trajectory.
From lemma 3.1 we need only to prove that for any $\xi\in U$
\[
E\xi^{2}-c^{2}+\lambda_{1}(X_{0}^{\xi}-y)+\lambda_{2}(E\xi-c)\geq E(\xi^{\ast
})^{2}-c^{2}+\lambda_{1}(X_{0}^{\xi^{\ast}}-y)+\lambda_{2}(E\xi^{\ast}-c),
\]
i.e., to prove%
\[
E\xi^{2}-E(\xi^{\ast})^{2}+\lambda_{1}(X_{0}^{\xi}-X_{0}^{\xi^{\ast}}%
)+\lambda_{2}E(\xi-\xi^{\ast})\geq0.
\]

Set%
\begin{align*}
\hat{\xi}  &  =\xi-\xi^{\ast},\\
f_{1}(x,z,t)  &  =f(X^{\ast}(t)+x,Z^{\ast}(t)+z,t)-f(X^{\ast}(t),Z^{\ast
}(t),t),\\
f_{2}(x,z,t)  &  =f_{X}(X^{\ast}(t),Z^{\ast}(t),t)x+f_{Z}(X^{\ast}(t),Z^{\ast
}(t),t)z.
\end{align*}

Consider the following equation%
\[
\left\{
\begin{array}
[c]{rl}%
-d(X(t)-X^{\ast}(t)) & =[f(X(t),Z(t),t)-f(X^{\ast}(t),Z^{\ast}%
(t),t)]dt-(Z(t)-Z^{\ast}(t))^{\prime}dW(t),\\
& =[f_{1}(X(t)-X^{\ast}(t),Z(t)-Z^{\ast}(t),t)dt-(Z(t)-Z^{\ast}(t))^{\prime
}dW(t),\\
X(T)-X^{\ast}(T) & =\hat{\xi}.
\end{array}
\right.
\]

By Assumption (H3),
\[
f_{1}(x,z,t)\geq f_{2}(x,z,t)\qquad\forall x,z,\quad dP\otimes dt-a.s.
\]

Hence applying the comparison theorem for BSDEs, we obtain $X(t)-X^{\ast
}(t)\geq\delta X(t),$ $\forall t$ $P-a.s.,$ where $\delta X(\cdot)$ is the
solution of (3.2).

Using the following inequality%
\[
(\xi^{\ast})^{2}-\xi^{2}\leq-2\xi^{\ast}(\xi-\xi^{\ast})
\]
and (3.6), we have%
\[%
\begin{array}
[c]{ll}
& E\xi^{2}-E(\xi^{\ast})^{2}+\lambda_{1}(X_{0}^{\xi}-X_{0}^{\xi^{\ast}%
})+\lambda_{2}E(\xi-\xi^{\ast})\\
\geq & 2E[\xi^{\ast}(\xi-\xi^{\ast})]+\lambda_{1}\delta X(0)+\lambda_{2}%
E(\xi-\xi^{\ast})\\
\geq & 2E(\xi^{\ast}\hat{\xi})+\lambda_{1}\delta X(0)+\lambda_{2}E\hat{\xi}\\
\geq & E[(2\xi^{\ast}+\lambda_{1}q(T)+\lambda_{2})\hat{\xi}].
\end{array}
\]

Since (3.4) implies%
\[
E[(2\xi^{\ast}+\lambda_{1}q(T)+\lambda_{2})\hat{\xi}]\geq0,
\]
we obtain the result. The proof is complete. $\blacksquare$

\section{Existence of the optimal solution}

In this section, we prove that there exists a unique optimal solution for the
optimization problem (2.7). We also show that the optimal solution can be
obtained by solving a FBSDE with constraints.

\begin{theorem}
Suppose that (H1)-(H6) hold. Then there exists a unique $\xi^{\ast}\in
L^{2}(\Omega,\mathcal{F}_{T},P)$ which attains the minimum of the problem (2.7).
\end{theorem}

\textbf{Proof.} The uniqueness is due to the strict convexity of the
functional
\[
\xi\mapsto J(\xi),\quad\xi\in U.
\]

As for the existence, consider the set given by%
\[
\mathcal{B}=\{\xi\in\mathcal{N}(y);\quad J(\xi)\leq C\}
\]
where $C>0$ is a constant. It is clear that, for each constant $C$,
$\mathcal{B}$ is bounded, closed and convex. Hence $\mathcal{B}$ is weakly
compact and by classical results of convex analysis \cite{br}, we need only to
show that $J$ is weakly lower-semicontinuous. Since $J$ is convex and strongly
lower-semicontinuous (in fact, it is strongly continuous [6, 7]), it follows
that $J$ is lower-semicontinuous for the weak convergence \cite{br}.

Thus the minimum of the problem (2.7) is attained (refer to Corollary 3.20 in
\cite{br}). The proof is complete. $\blacksquare$

\begin{corollary}
We assume (H1)-(H6). Then there exist constants $\lambda_{1}>0$ and
$\lambda_{2}\in R$ such that the optimal $\xi^{\ast}$ has the form%
\begin{equation}
\xi^{\ast}=\frac{1}{2}(-\lambda_{2}-\lambda_{1}q(T))^{+}. \tag{4.1}%
\end{equation}

\end{corollary}

This is a direct consequence of Theorem 3.3 and Theorem 4.1. The proof is omitted.

Let $(X^{\ast}(\cdot),Z^{\ast}(\cdot))$ be the optimal wealth process and
portfolio associated with $\xi^{\ast}$ for problem (2.7).

\begin{theorem}
Suppose that (H1)-(H6) hold. Then there exist a positive number $\lambda_{1}$
and $\lambda_{2}\in R$ such that the following FBSDE%
\begin{equation}
\left\{
\begin{array}
[c]{l}%
dq(t)=q(t)[f_{X}(X(t),Z(t),t)dt+f_{Z}(X(t),Z(t),t)^{\prime}dW(t)],\\
q(0)=1,\\
-dX(t)=f(X(t),Z(t),t)dt-Z(t)^{\prime}dW(t),\\
X(T)=\frac{1}{2}(-\lambda_{2}-\lambda_{1}q(T))^{+}%
\end{array}
\right.  \tag{4.2}%
\end{equation}
with constraints
\begin{equation}
EX(T)=c\quad\text{and}\quad X(0)=y \tag{4.3}%
\end{equation}
has a unique solution $(q(\cdot),X(\cdot),Z(\cdot)).$ Furthermore, we have
$(X(\cdot),Z(\cdot))=(X^{\ast}(\cdot),Z^{\ast}(\cdot))$ and $X(T)=\xi^{\ast}$.
\end{theorem}

\textbf{Proof.} Note that (4.1) is equivalent to (3.4). Then it is easy to
check that the solution of FBSDE (4.2) with (4.3) is just the optimal solution
of problem (2.7) by Theorem 3.3 and Theorem 4.1. The proof is complete.
$\blacksquare$

Finally, we show that the smoothness condition, i.e., Assumption (H1) may not
hold for the following examples:

\begin{example}
Suppose that taxes must be paid on the gains which are made on the risky
securities. The wealth process $X$ is governed by
\begin{equation}
\left\{
\begin{array}
[c]{l}%
-dX(t)=-[r(t)X(t)+\pi(t)^{\prime}\sigma(t)\theta(t)-\alpha(\pi(t)^{\prime
}\sigma(t)\theta(t))^{+}]dt-\pi(t)^{\prime}\sigma(t)dW(t),\\
X(0)=x.
\end{array}
\right.  \tag{4.4}%
\end{equation}

\end{example}

\begin{example}
Suppose that the borrowing interest rate $R(t)\geq r(t)$. In this case, the
wealth process $X$ satisfies%
\begin{equation}
\left\{
\begin{array}
[c]{l}%
-dX(t)=-[r(t)X(t)+\pi(t)^{\prime}\sigma(t)\theta(t)-(R(t)-r(t))(X(t)-%
{\displaystyle\sum\limits_{i=1}^{d}}
\pi_{i}(t))^{-}]dt-\pi(t)^{\prime}\sigma(t)dW(t),\\
X(0)=x.
\end{array}
\right.  \tag{4.5}%
\end{equation}

\end{example}

But in this case, we can still prove that (3.4) is a sufficient condition for
optimality. To this end, we need an additional assumption:

(H1)' $f$ is uniformly Lipschitz with respect to $(X,Z)$.

Let $\xi^{\ast}\in U$ and $(X^{\ast}(\cdot),Z^{\ast}(\cdot))$ be the
corresponding trajectory.

\begin{theorem}
Suppose that (H1)' and (H2)-(H6) hold. If there exist constants $\lambda
_{1}>0$ and $\lambda_{2}\in R$ such that (3.4) with $X_{0}^{\xi^{\ast}}=y$ is
satisfied or equivalently, (4.2) with (4.3) has a solution $(q(\cdot),X^{\ast
}(\cdot),Z^{\ast}(\cdot))$, then $\xi^{\ast}=X^{\ast}(T)$ is an optimal
terminal wealth for problem (2.7).
\end{theorem}

\textbf{Proof.} We should only use subdifferentials instead of differentials
in the second part proof of Theorem 3.3. Note that now $f_{X}$ (resp. $f_{Z}%
$)denotes a predictable process belonging $dP\otimes dt$ almost surely to
$\partial f(X^{\ast}(t),Z^{\ast}(t),t)$, where $\partial f$ is the
subdifferential of $f$ with respect to $X$ (resp. $Z$).

The proof is complete. $\blacksquare$

\section{Concluding remarks}

This paper investigates the continuous-time mean-variance portfolio selection
model with nonlinear wealth equation and bankruptcy prohibition. A stochastic
maximum principle is established via the dual method and terminal perturbation
technique. Under the smoothness conditions on the coefficients (Assumption
(H1)), we prove that the established stochastic maximum principle is not only
a necessary but also a sufficient condition for the optimal terminal wealth.
Then the optimal wealth and portfolio strategy, i.e., the solution of the
FBSDE (4.2) can be computed by the PDE approach of Ma, Protter and Yong
\cite{ma1}, the probability method of Hu and Peng \cite{hu} or numerical
methods (see also \cite{ma2} for systematical investigation). If the
smoothness assumption does not hold, we only obtain a sufficient condition,
i.e., Theorem 4.6. In this case, the main difficulty lies in the fact that the
corresponding FBSDE (4.2) may have discontinuous coefficients. We emphasize
that it remains an interesting open problem to solve FBSDEs with discontinuous
coefficients. But as shown in Theorem 4.6, our method in this paper can be
used to derive the existence of solutions for FBSDE (4.2). Another important
point to note here is that the existing results in the utility framework can't
cover the mean-variance model at all since the usual assumptions imposed on
utility functions are different from those on the mean-variance models.

\vskip0.5cm

{\large Appendix A.}

\vskip0.2cm

\textbf{Feasibility analysis.}

For a given initial investment $x>0$ and $c>0$, if there exists a portfolio
$\pi(\cdot)\in\mathcal{A}(x)$, the initial investment $x$ is called
admissible. Our aim is to compute the minimal admissible initial investment
which is denoted by $\bar{x}$. If $\bar{x}\leq y$ (resp. $\bar{x}<y)$, the
optimization problem (2.4) and (2.7) are feasible (resp. the slater condition holds).

Using similar analysis as in section 2, we can obtain $\bar{x}$ by solving the
following optimization problem:
\[
\bar{x}=\underset{\xi\in U}{\inf}X_{0}^{\xi},
\]%
\[
\text{subject\thinspace\thinspace to \ }E\xi=c.
\]

For $\lambda\in R$, define%
\[
\varphi(\lambda)=\underset{\xi\in U}{\inf}[X_{0}^{\xi}+\lambda E(\xi-c)].
\]

By the classical results of duality theory \cite{lu}, we have%
\[
\bar{x}=\underset{\lambda\in R}{\max}\varphi(\lambda).
\]

\vskip0.5cm

{\large Appendix B.}

\vskip0.2cm

\textbf{Proof of Lemma 3.2}. From (2.6) and (3.2), we have%
\[
\left\{
\begin{array}
[c]{ll}%
-d\tilde{X}^{\rho}(t) & =\rho^{-1}[f(X^{\rho}(t),Z^{\rho}(t),t)-f(X^{\ast
}(t),Z^{\ast}(t),t)-\rho f_{X}(X^{\ast}(t),Z^{\ast}(t),t)\delta X(t)\\
& -f_{Z}^{^{\prime}}(X^{\ast}(t),Z^{\ast}(t),t)\delta Z(t)]dt-\tilde{Z}^{\rho
}(t)^{\prime}dW(t),\\
\tilde{X}^{\rho}(T) & =0.
\end{array}
\right.
\]
Let%
\begin{align*}
A^{\rho}(t)  &  =\int_{0}^{1}f_{X}(X^{\ast}(t)+\lambda\rho(\delta
X(t)+\tilde{X}^{\rho}\left(  t\right)  ),Z^{\ast}(t)+\lambda\rho(\delta
Z(t)+\tilde{Z}^{\rho}(t)),t)d\lambda,\\
B^{\rho}(t)  &  =\int_{0}^{1}f_{Z}(X^{\ast}(t)+\lambda\rho(\delta
X(t)+\tilde{X}^{\rho}\left(  t\right)  ),Z^{\ast}(t)+\lambda\rho(\delta
Z(t)+\tilde{Z}^{\rho}(t)),t)d\lambda,\\
C^{\rho}(t)  &  =[A^{\rho}(t)-f_{X}(X^{\ast}(t),Z^{\ast}(t),t)]\delta
X(t)+[B^{\rho}(t)-f_{Z}(X^{\ast}(t),Z^{\ast}(t),t)]\delta Z(t).
\end{align*}
Thus%
\[
\left\{
\begin{array}
[c]{l}%
-d\tilde{X}^{\rho}\left(  t\right)  =(A^{\rho}(t)\cdot\tilde{X}^{\rho}\left(
t\right)  +B^{\rho}(t)\cdot\tilde{Z}^{\rho}(t)+C^{\rho}(t))dt-\tilde{Z}^{\rho
}(t)^{\prime}dW(t),\\
\tilde{X}^{\rho}(T)=0
\end{array}
\right.
\]
Using It\^{o}'s formula to $\mid\tilde{X}^{\rho}\left(  t\right)  \mid^{2}$ we
get%
\begin{align*}
E  &  \mid\tilde{X}^{\rho}\left(  t\right)  \mid^{2}+E\int_{t}^{T}\mid
\tilde{Z}^{\rho}(s)\mid^{2}ds\\
\  &  =2E\int_{t}^{T}\tilde{X}^{\rho}(s)(A^{\rho}(s)\cdot\tilde{X}^{\rho
}(s)+B^{\rho}(s)\cdot\tilde{Z}^{\rho}(s)+C^{\rho}(s))ds\\
\  &  \leq KE\int_{t}^{T}\mid\tilde{X}^{\rho}(s)\mid^{2}ds+\frac{1}{2}%
E\int_{t}^{T}\mid\tilde{Z}^{\rho}(s)\mid^{2}ds+E\int_{t}^{T}\mid C^{\rho
}(s)\mid^{2}ds
\end{align*}

where $K$ is a constant. So%
\begin{align*}
E  &  \mid\tilde{X}^{\rho}\left(  t\right)  \mid^{2}+\frac{1}{2}E\int_{t}%
^{T}\mid\tilde{Z}^{\rho}(s)\mid^{2}ds\\
&  \leq KE\int_{t}^{T}\mid\tilde{X}^{\rho}(s)\mid^{2}ds+E\int_{t}^{T}\mid
C^{\rho}(s)\mid^{2}ds
\end{align*}

By the Lebesgue dominate convergence theorem, we have%
\[
\underset{\rho\rightarrow0}{\lim}E\int_{0}^{T}\mid C^{\rho}(t)\mid^{2}dt=0.
\]

Applying Grownwall's inequality, we obtain the result. $\blacksquare$

\vskip0.5cm

\textbf{Acknowledgement. }The author would like to thank Prof. Shige Peng and
Dr. Hanqing Jin for some useful conversations. Especially, Assumptions
(H5)-(H6) are due to helpful discussions with Prof. Xunyu Zhou.

\end{document}